\documentclass[prb,twocolumn,showpacs,preprintnumbers,amsmath,amssymb]{revtex4}
\usepackage{graphicx}
\usepackage{bm}
\usepackage{epsfig}

\begin{document}

\newcommand{\bec}{\begin{center}}
\newcommand{\ec}{\end{center}}
\newcommand{\be}{\begin{equation}}
\newcommand{\ee}{\end{equation}}
\newcommand{\beqn}{\begin{eqnarray}}
\newcommand{\eeqn}{\end{eqnarray}}
\newcommand{\bet}{\begin{table}}
\newcommand{\ent}{\end{table}}
\newcommand{\bib}{\bibitem}



\title{
Anisotropy driven ultrafast nanocluster burrowing  
}

\author{P. S\"ule} 
\affiliation{
  Research Institute for Technical Physics and Material Science,\\
Konkoly Thege u. 29-33, Budapest, Hungary,sule@mfa.kfki.hu,www.mfa.kfki.hu/$\sim$sule,\\
}


\date{\today}

\begin{abstract}
We explore the occurrence of 
low-energy and low-temperature transient cluster burrowing leading to intact cluster inclusions.
In particular, the anomalously fast (ballistic) Pt nanocluster implantation into Al and Ti substrates has been found by molecular dynamics
simulations using a tight-binding many-body potential with
the $1-5$ eV/atom low impact energy.
Similar behavior has also been found for many other cluster/substrate couples such as Cu/Al and
Ni/Ti, Co/Ti, etc. In particular, in Ni/Ti at already $\sim 0.5$ eV/atom impact energy burrowing takes place.
At this few eV/atom low impact energy regime instead of the expected stopping at the surface we find the propagation of the cluster
through a thin Al slab as thick as $\sim 50$ $\hbox{\AA}$ with a nearly constant speed ($\propto 1$ eV/atom).
Hence the cluster moves far beyond the range of the impact energy which
suggests that the mechanism of cluster burrowing
can not be explained simply by collisional cascade effects. 
In the couples with reversed succession (e.g. Ti/Pt, Al/Pt) no burrowing has been
found, the clusters remain on the surface (the asymmetry of burrowing).
We argue that cluster penetration occurs at few eV/atom impact energy when the cluster/substrate
interaction is size-mismatched and mass anisotropic atomically.


\end{abstract}
\maketitle

\section{Introduction}

 Cluster deposition on solid surfaces has been the subject of intense atomistic level studies in
the last decades \cite{clustersatsurfaces,Jensen,cluster,Kai,Zimmermann,Pratontep,Carroll}.
The reason of the considerable interest is due to that 
recent progress in this field offers a great technological potential
for the application of clusters in various areas, such
as the smoothening and cleaning of surfaces \cite{smooth}, the improvement of magnetic properties \cite{magnetic}, size-dependent catalytic activity \cite{catalytic},
thin film growth with a well-controlled grain size \cite{clustersatsurfaces,Jensen,cluster}
or the formation of cluster assembled nanowires \cite{Schmelzer}.

  Cluster-surface interaction as well as the mobility of clusters
on surfaces have also been intensively studied \cite{cluster,Massobrio,burrowing,Stepanyuk,CoonAu}.
Transient mobility (TM) of clusters has been known for a while on surfaces \cite{Jensen,Massobrio,Vasco,Luedtke} but not in the bulk.
Hereby we report on the possible occurrence of TM of various clusters perpendicular to the surface
leading to transient cluster burrowing into various substrates.
Until now only
slow cluster burrowing with soft landing has been studied in few systems which takes place at least in a nanosecond time scale well above room temperature \cite{burrowing,Stepanyuk,CoonAu}. 
The high energy ion-bombardment induced burrowing of Pt nanoparticles
into SiO$_2$ has also been reported recently \cite{Hu}.

 Cluster impact phenomena have been intensively studied by computer atomistic
simulations to understand the processes which govern
the creation of nanoscale structures on the surface \cite{clustersatsurfaces,Jensen,cluster,Kai,Pratontep,Carroll,burrowing,Stepanyuk,CoonAu}.
However, at best of our knowledge no low-energy (few eV/atom or less) and low temperature cluster beam deposition with transient implantation rate has been reported
until now in which the cluster remains intact in the bulk.
Implantation of clusters with few tens of $\hbox{\AA}$ depth usually requires at least few tens of eV impact energy \cite{Jensen} or even higher energy is required for deeper penetration depths \cite{Pratontep,Carroll}.
Cluster implantation at much lower energies could offer great technological
possibilites for the production of nanostructures due to the much less destructive conditions.

  According to our recent findings that single atomic deposition in Pt/Al(111) \cite{Sule_JCP} and interfacial intermixing in Pt/Ti bilayer \cite{Sule_JAP07} 
are anomalous in many respect, e.g. transient inter-layer atomic transport rates have been found,
it could be that cluster transport in systems with the same atomic constituents could also be
anomalously fast.
Moreover, it has also been shown recently that in mass-anisotropic bilayers the enhancement 
of interfacial broadening and intermixing takes place \cite{Sule_PRB05,Sule_SUCI} and it could also be 
that mass-anisotropy does influence the propagation of deposited clusters.

 In this article 
 we report on the occurrence of ultrafast cluster transport
(burrowing) in Al (and in Ti, Ag and in Au)
bulk upon low-energy impact ($<$ few eV/atom) at $\sim 0$ K predicted
by atomistic molecular dynamics (MD) simulations.
The employed interpolated crosspotentials have been fitted to
{\em ab initio} calculations.
We show that the Pt cluster initialized with few eV/atom impact energy can move ballistically in the bulk of Al until few tens of ps.
In other deposition events with various cluster/substrate couples we find 
similar 
behavior when the cluster to substrate interaction is atomically size mismatched
and/or mass anisotropic in such a way that the cluster is composed of
heavier and smaller atoms.

\section{The simulation method}

\subsection{General properties}

 Classical constant volume tight-binding molecular dynamics simulations \cite{CR} were used to simulate 
soft landing and low-energy cluster impact of various nanoclusters on (111) surfaces of Al, Ti, Ag and Au at $\sim 0$ K
initial temperature
using the PARCAS code \cite{Kai,burrowing,PARCAS}.
The MD code has widely been used for the study of various atomic transport phenomena
in the last few years \cite{Kai,Sule_JCP,Sule_JAP07,Sule_PRB05,burrowing,PARCAS}.
Although we carry out simulations at $\sim 0$ K, we find a substantial local heating
up in a local surface region of Al, hence the correct dissipation of the
emerged heat should be handled using temperature control.
A variable timestep
and the Berendsen temperature control is used at the cell border \cite{Berendsen}.
The simulation uses the Gear's predictor-corrector algorithm to calculate
atomic trajectories \cite{PARCAS}.
The maximum time step of $0.05$ fs is used during the operation of the multiple time step algorithm.
The system couples to a heat bath via the damping constant 
to maintain constant temperature conditions and the thermal equilibrium of the entire
system \cite{Berendsen}.
The time constant for temperature control is chosen to be $\tau=70$ fs, where $\tau$ is a characteristic relaxation time to be adjusted \cite{Berendsen}.
The Berendsen temperature control has successfully been used 
for nonequilibrium systems, such as occur during ion-bombardment of various
materials \cite{Kai,burrowing,Sule_JAP07,Sule_PRB05,PARCAS}.

  For simulating deposition it is appropriate to use temperature control
at the cell borders.
This is because it is physically correct that potential energy becomes
kinetic energy on impact, i.e. heats the lattice. This heating should
be allowed to dissipate naturally, which means temperature control should not
be used at the impact point.
Periodic boundary conditions are imposed laterarily.
The observed anomalous transport processes are also observed without
periodic boundary conditions and Berendsen temperature control.
Further details are given in \cite{PARCAS} and details specific to the current and similar systems in recent
communications \cite{Sule_JCP,Sule_JAP07,Sule_PRB05}.

  The top of the simulation cell is left free (the free surface) for
the deposition of Pt atoms.
The bottom layers
are held fixed in order to avoid the rotation of the cell.
Since the z direction is open, rotation could start around the z axis.
The bottom layer fixation is also required to prevent
the translation of the cell.
In few cases, however, we do not fix bottom layers, when the penetration
of the deposited cluster through thin substrate films
has been demonstrated.
No rotation of the cell has been found in these cases.

  The size of the typical simulation cell for the Pt/Al system is $80 \times 80 \times 42$ $\hbox{\AA}^3$ including
 $16128$ substrate atoms (with a fcc lattice,
$15$ active MLs are supported on $3$ fixed bottom monolayers (MLs)).
We also tried substrates with different lateral sizes ($80-160$ $\hbox{\AA}^3$) and slab thickness ($42-100$ $\hbox{\AA}^3$) up to $\sim 150000$ number of atoms.
Other cluster/substrates couples 
(Pt/Ti, Co/Au, Cu/Al, Ni/Ti) together with reversed succession (Ti/Pt, Ti/Ni, etc.)
have also been constructed with similar system size.
We tested cluster impact on much larger substrates and
find no dependence of the anomalous atomic transport properties of the deposited atoms on the
finite size of the simulation cell.
The cubic cluster includes roughly $\sim 1500$ atoms (the length of cube edges: $\sim 30$ $\hbox{\AA}$).
We also tested the penetration of various clusters in different substrates
with similar lateral size and slab thickness and cluster size.
The kinetic energy of the deposited particles is
in the range of $\sim 1-10$ eV/atom (the velocity of the clusters normal to the surface is $v \approx 20-200$ $\hbox{\AA}$/ps).
The cluster is initialized at $\sim 8$ $\hbox{\AA}$ above the (111) surface of the substrate.

\begin{table}
\begin{center}
\caption
{
The parameters used in the interpolated tight binding potential (TB-SMA) given in Eqs
. (1)-(2) \cite{CR}
The parameters of the crosspotential have been obtained
as follows using an interpolation scheme \cite{Sule_PRB05}:
For the preexponentials $\xi$ and $A$ we used the
harmonic mean $A_{AlPt}=(A_{Al} \times A_{Pt})^{1/2}$
for $q$ and $p$ we use the geometrical averages:
$q_{AlPt}=(q_{Al} + q_{Pt})/2$. The first neighbor distance of the
Al-Pt potential is given also as a geometrical mean of $r_0=(r_0^{Pt}+r_0^{Al})/2$.
In order to get better agreement between {\em ab initio}
and semiempirical potentials
parameters $\xi$ and $r_0$ have been optimized.
In the case of Co/Au parameter $q$ has also been fitted.
}
{\large
\begin{tabular}{cccccc}
& $\xi$ & q & A & p & $r_0$  \\
\hline \hline
 Al-Pt & 2.7  & 3.258 & 0.191  & 9.612 & 3.0   \\
 Ti-Pt & 4.2  & 2.822 & 0.149  & 11.015 & 2.87   \\
 Ni-Ti & 0.8 & 1.416 & 0.052  & 14.209 & 2.72  \\  
 Cu-Al & 2.1 & 2.397 & 0.102  & 9.785  & 2.85  \\ 
 Co-Au & 2.5 & 2.7   & 0.140  & 10.917 & 2.85      \\
\hline \hline
\end{tabular}
}
\end{center}
\label{table}
\end{table}

\subsection{The interatomic potentials}

  We use the many-body Cleri-Rosato (CR) parametriz\-ation of the tight-binding 
second-moment approximation (TB-SMA) interaction potential to describe interatomic 
interactions \cite{CR}.
\begin{figure}[hbtp]
\begin{center}
\includegraphics*[height=3.5cm,width=4cm,angle=0.]{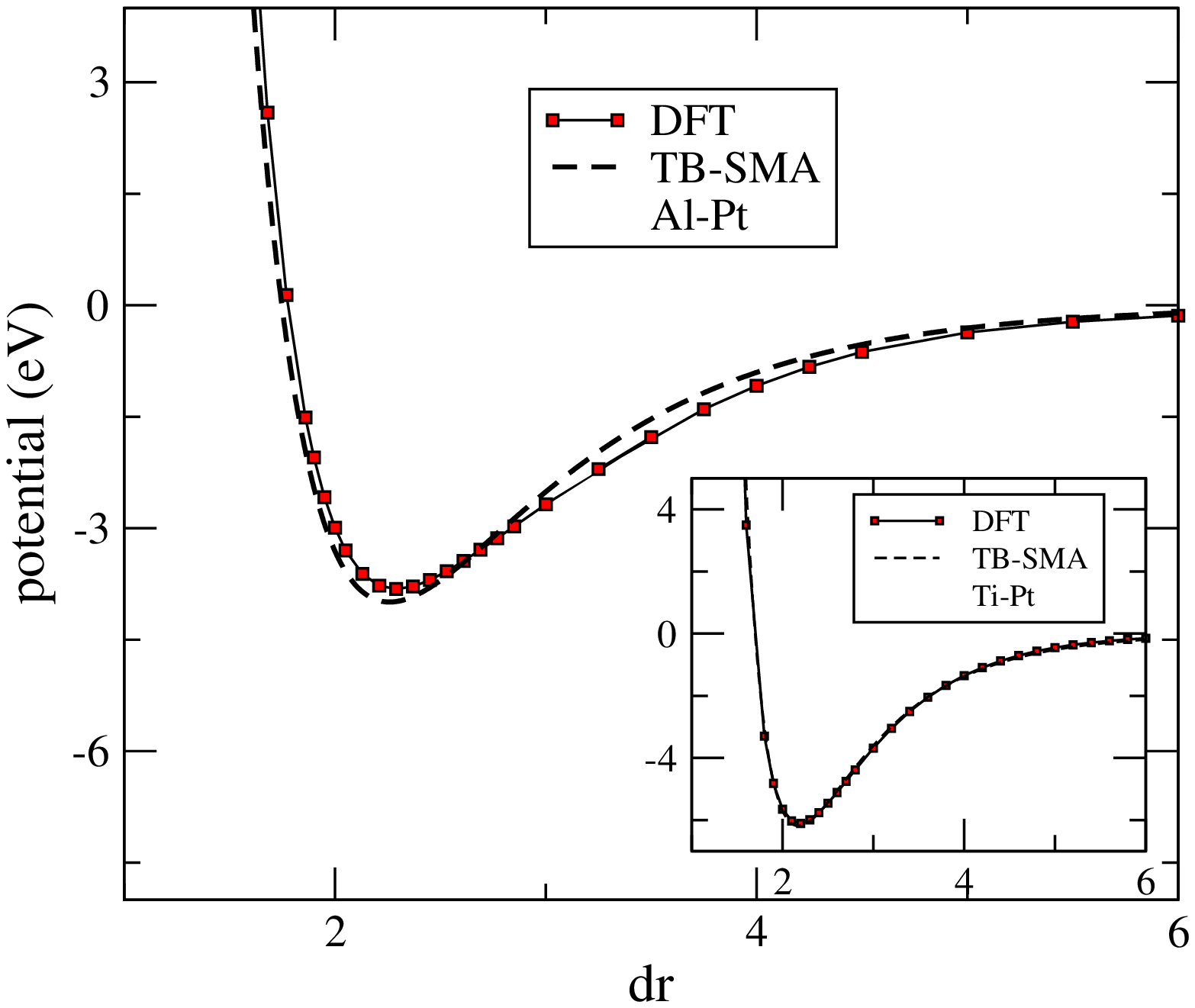}
\includegraphics*[height=3.5cm,width=4cm,angle=0.]{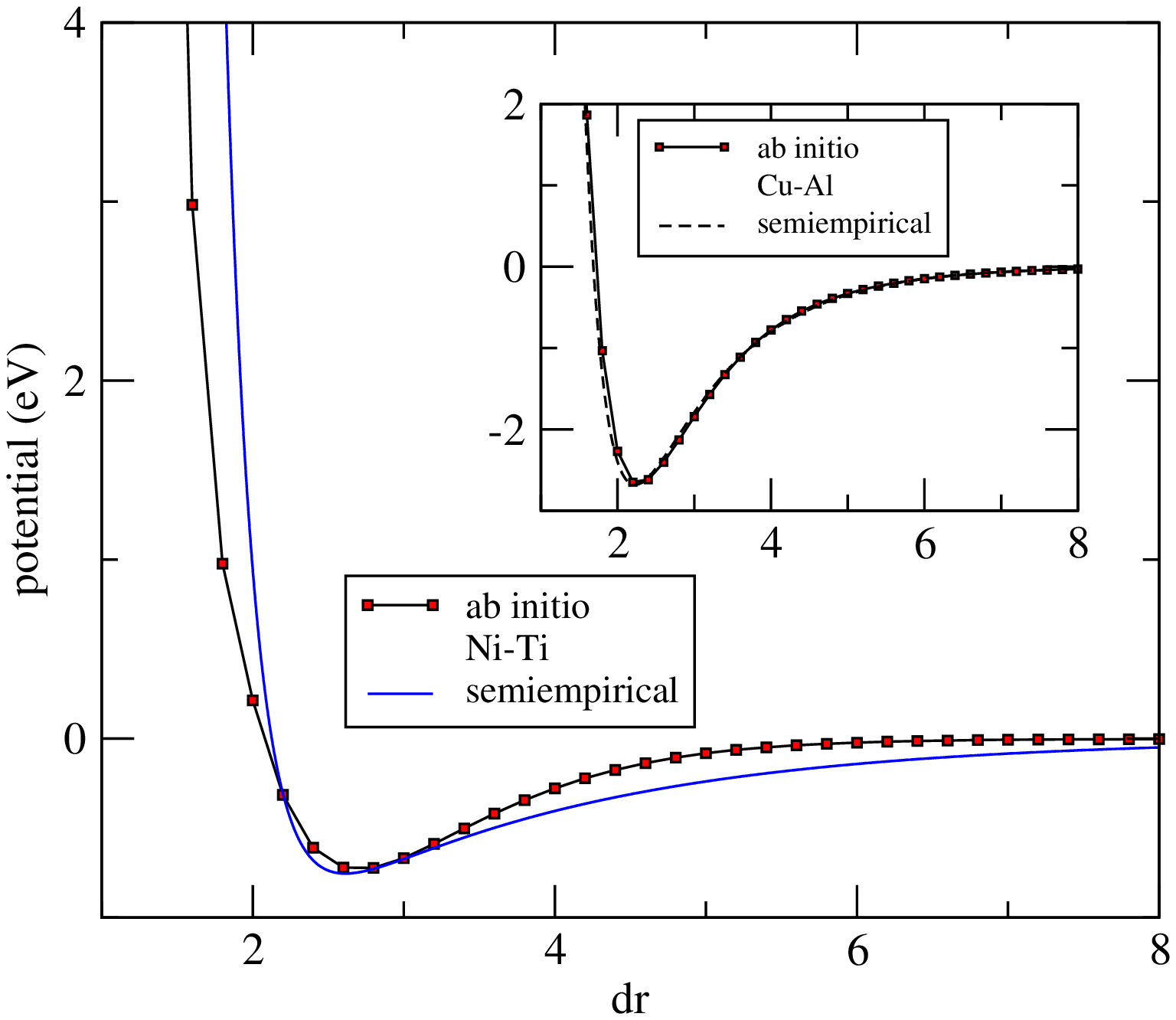}
\includegraphics*[height=3.5cm,width=4cm,angle=0.]{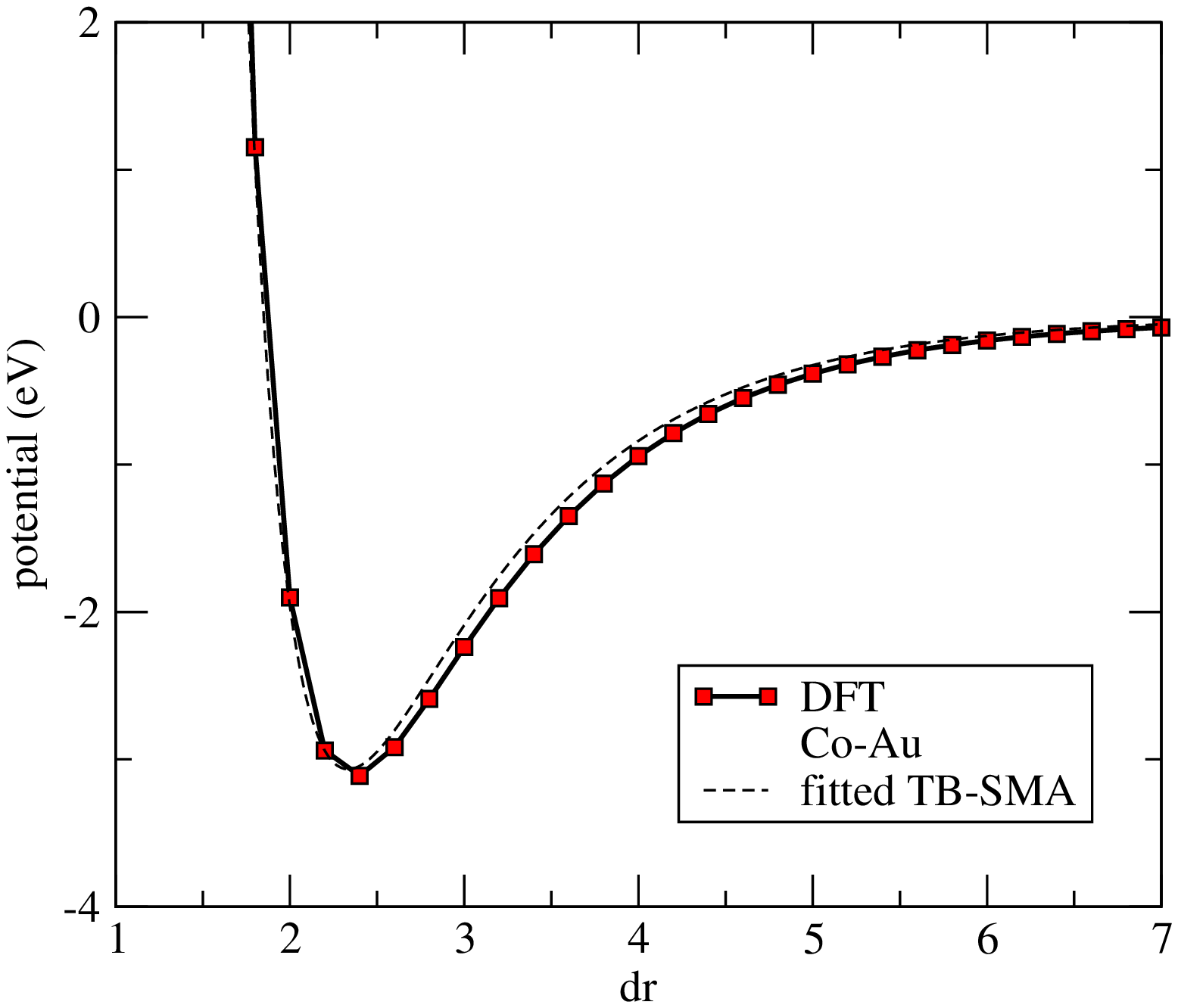}

\caption[]{
The crosspotential energy (eV) for the dimers Al-Pt, Ti-Pt (inset Fig. 1a), Ni-Ti, Cu-Al (inset Fig. 2a) and Co/Au as a function of
the interatomic distance ($\hbox{\AA}$) obtained by
the {\em ab initio} PBE/DFT method. For comparison the fitted interpolated semiempirical pote
ntials
(TB-SMA) are also shown.
The fit procedure has been carried out by varying parameters
$\xi$ and $r_0$ in order to get the best macth of the potential energy
curves with the {\em ab initio} one.

}
\label{potential}
\end{center}
\end{figure}

 Within the TB-SMA, the band energy
(the attractive part of the potential) reads,
\be
 E_b^i=-\biggm[ \sum_{j, r_{ij} < r_c} \xi^2 exp \biggm[-2q \biggm(\frac{r_{ij}}{r_0}
-1 \biggm) \biggm] \biggm]^{1/2},
\ee
where $r_c$ is the cutoff radius of the interaction and $r_0$ is the first neighbor distance (atomic size parameter).

The repulsive term is a Born-Mayer type phenomenological core-repulsion term:
\be
E_r^i=A \sum_{j, r_{ij}<r_c} exp \biggm[-p \biggm(\frac{r_{ij}}{r_0}-1 \biggm) \biggm].
\ee
The parameters ($\xi, q, A, p, r_0$) are fitted to experimental values of the cohesive energy,
the lattice parameter, the bulk modulus and the elastic constants $c_{11}$, $c_{12}$ and $c_{44}$ \cite{CR}
and which are given in Table 1.
The summation over $j$ is extended up to fifth neighbors for fcc structures \cite{CR}.
The cutoff radius $r_c$ is taken as the third neighbor distance for all the interactions.
We tested the Al-Al and the Al-Pt potential at cutoff radius with larger neighbor distances and
found no considerable change in the results.
This type of a potential gives a very good description of lattice vacancies, including migration
properties and a reasonable description of solid surfaces and melting \cite{CR}.
We also tested the Al-Al and the Al-Pt potential at cutoff radius with larger neighbor distances and found no considerable change in the results \cite{Sule_SUCI}.  
The cutoff radius is taken as the third-forth neighbor distance ($r_c \approx 10-15$ $\hbox{\AA}$) for all the interactions which we find sufficiently large enough.  

  For the crosspotential of substrate atoms and Pt we employ an interpolation scheme \cite{Sule_PRB05}
using
the geometrical mean of the elemental energy constants and the harmonic mean for the screening
length are taken as in refs. \cite{Sule_PRB05}.
The CR elemental potentials and the interpolation scheme for heteronuclear interactions
have widely been used for MD simulations \cite{Stepanyuk,Baletto,Sule_PRB05}.
Recently a CR interpolated crosspotential has also successfully been used for Ti/Pt in agreement with
our experimental results \cite{Sule_JAP07}.
The scaling factor $r_0$ (the heteronuclear first neighbor distance) is calculated as the average of the elemental first neighbor distances.
The AlPt potential has been fitted to the
measured effective heat of mixing in the cubic AlPt ($\Delta H \approx -100$ kJ/mol) \cite{Sule_PRB05}.
The melting temperature of $1870$ K \cite{Waal} is reproduced by our Cleri-Rosato
crosspotential within the range of $1800 \pm 100$ K.
In order to adjust $\Delta H$ in the Al-Pt potential (which is proportional
to the strength of the interaction and to the heat of alloying in the AlPt alloy)
the preexponential parameter $\xi$ in Eq (3) is set to $\xi \approx 3.0$ \cite{Sule_SUCI}.
Further details are given in \cite{PARCAS} and details specific to the current system in recent
communications \cite{Sule_JAP07,Sule_PRB05,Sule_SUCI}.

 In order to check the accuracy of the employed interpolated crosspotentials,
 the crosspotential energy has also been calculated for few of heteronuclear pairs (Al-Pt, Ti-Pt, Ni-Ti, Cu-Al, Co-Au)
  using {\em ab initio} local spin density functional calculations \cite{G03} together
with a quadratic convergence self-consistent field method.
The G03 code is well suited for molecular calculations, hence
it can be used for checking pair-potentials.
The interatomic potential $V(dr)$ between two atoms
is defined as the difference of total energy at an interatomic separation $dr$
and the total energy of the isolated atoms
\be
 V(dr)=E(dr)-E(\infty).
\ee
The Kohn-Sham equations (based on DFT) \cite{KS} are solved in an atom centered Gaussian basis set and the core electrons
are described by effective core potentials
(using the LANL2DZ basis set) \cite{basis}
and
we used the Perwed-Burke-Ernzerhof (PBE) gradient corrected exchange-correlation potential \cite{PBE}.
First principles calculations based on
density functional theory (DFT) have been applied in various fields
in the last few years \cite{DFT_Sule}.
The obtained profiles are plotted in Fig. \ref{potential}
together with our interpolated semiempirical many-body TB-SMA potentials for the various dimers.
Our interpolated potential is partly fitted to the {\em ab initio} one:
parameters $\xi$ and $r_0$ have been varied to get the best matching.
We find that our fitted interpolated TB-SMA potentials
match reasonably well the {\em ab initio} one hence we are convinced
that the TB-SMA model accurately describes the heteronuclear
interaction in the Al-Pt dimer.
Even if we use the interpolated TB-SMA without fitting, we get
nearly the
same results
\begin{figure}[hbtp]
\begin{center}
\includegraphics*[height=3.cm,width=4.3cm,angle=0.]{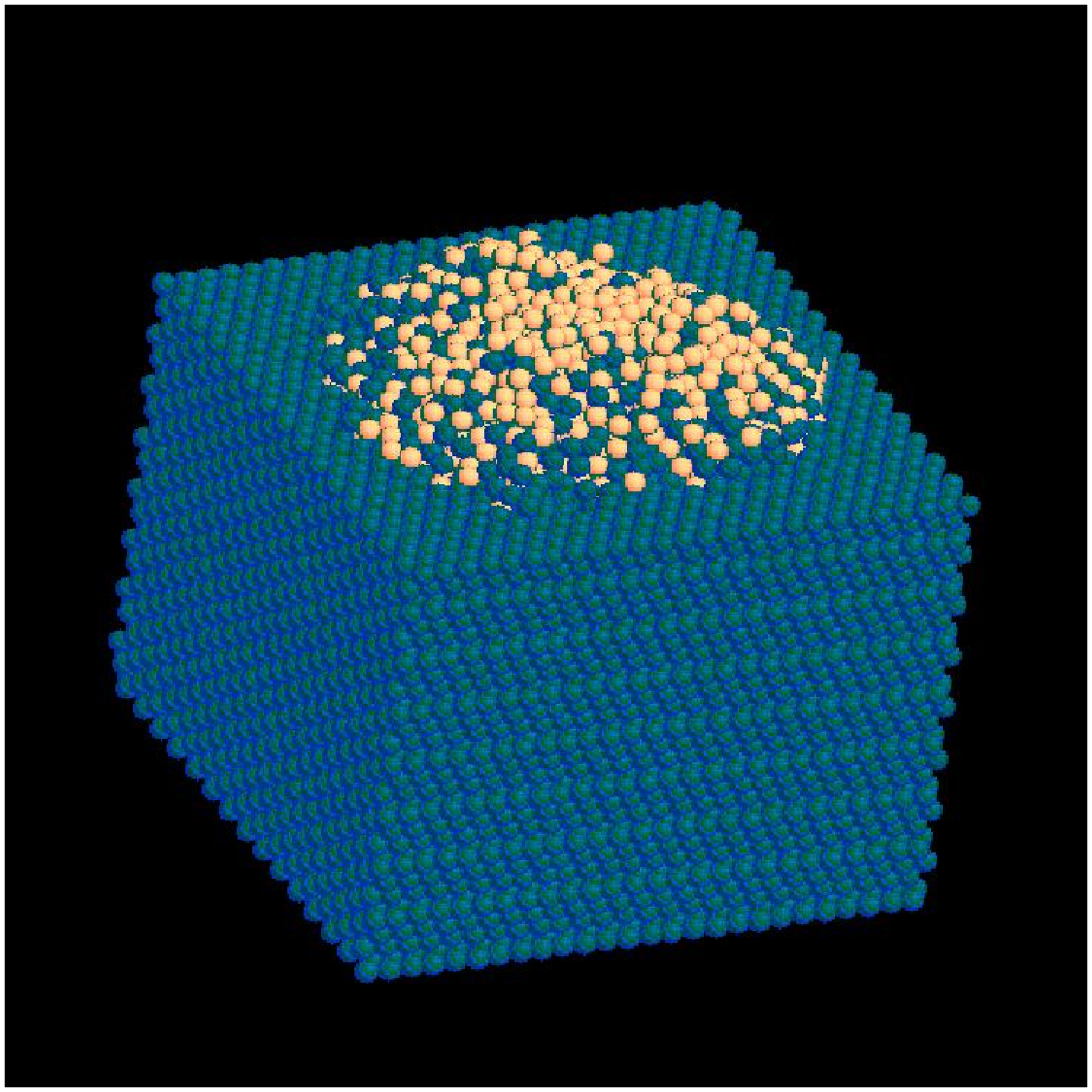}
\includegraphics*[height=3.cm,width=4.2cm,angle=0.]{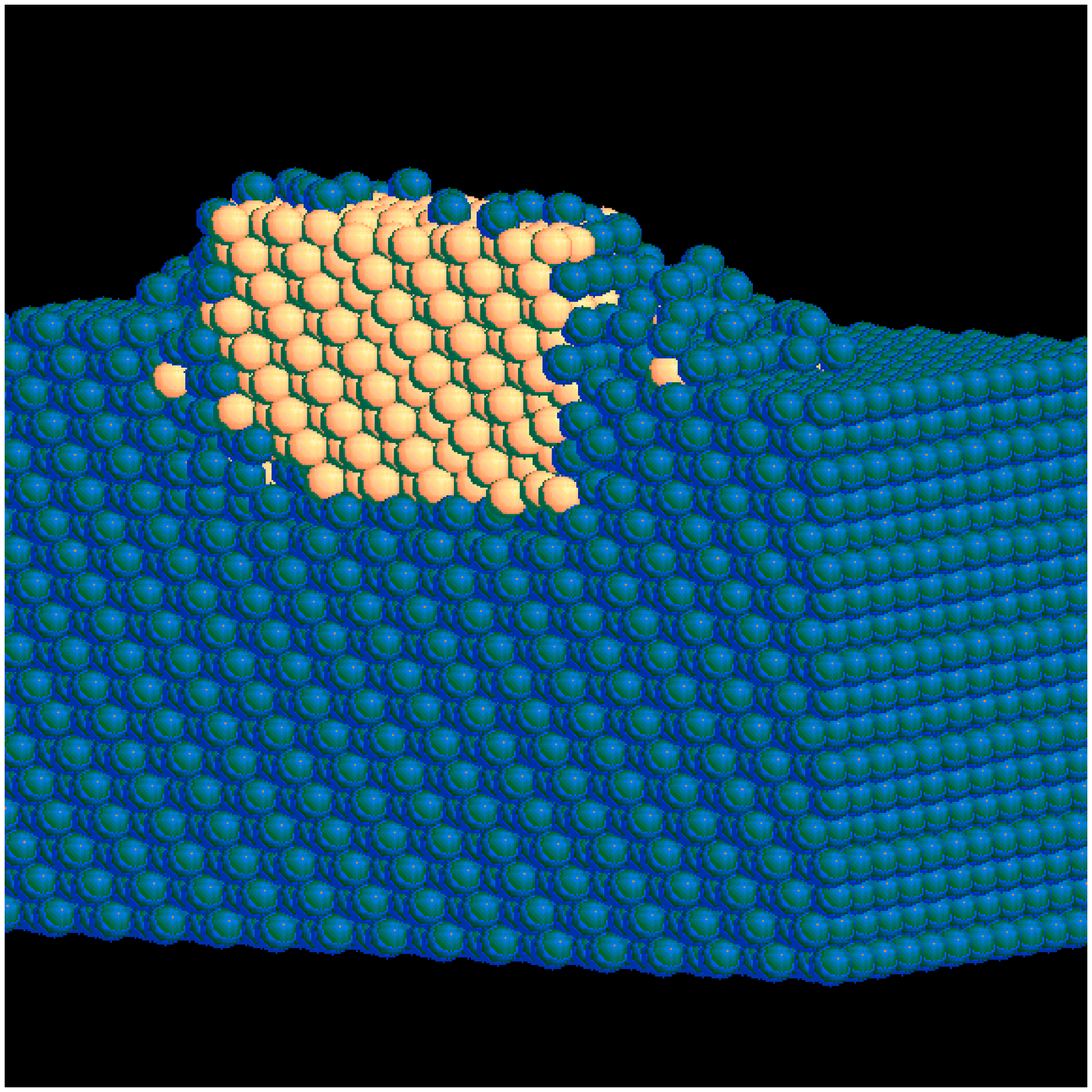}
\includegraphics*[height=3.cm,width=4.3cm,angle=0.]{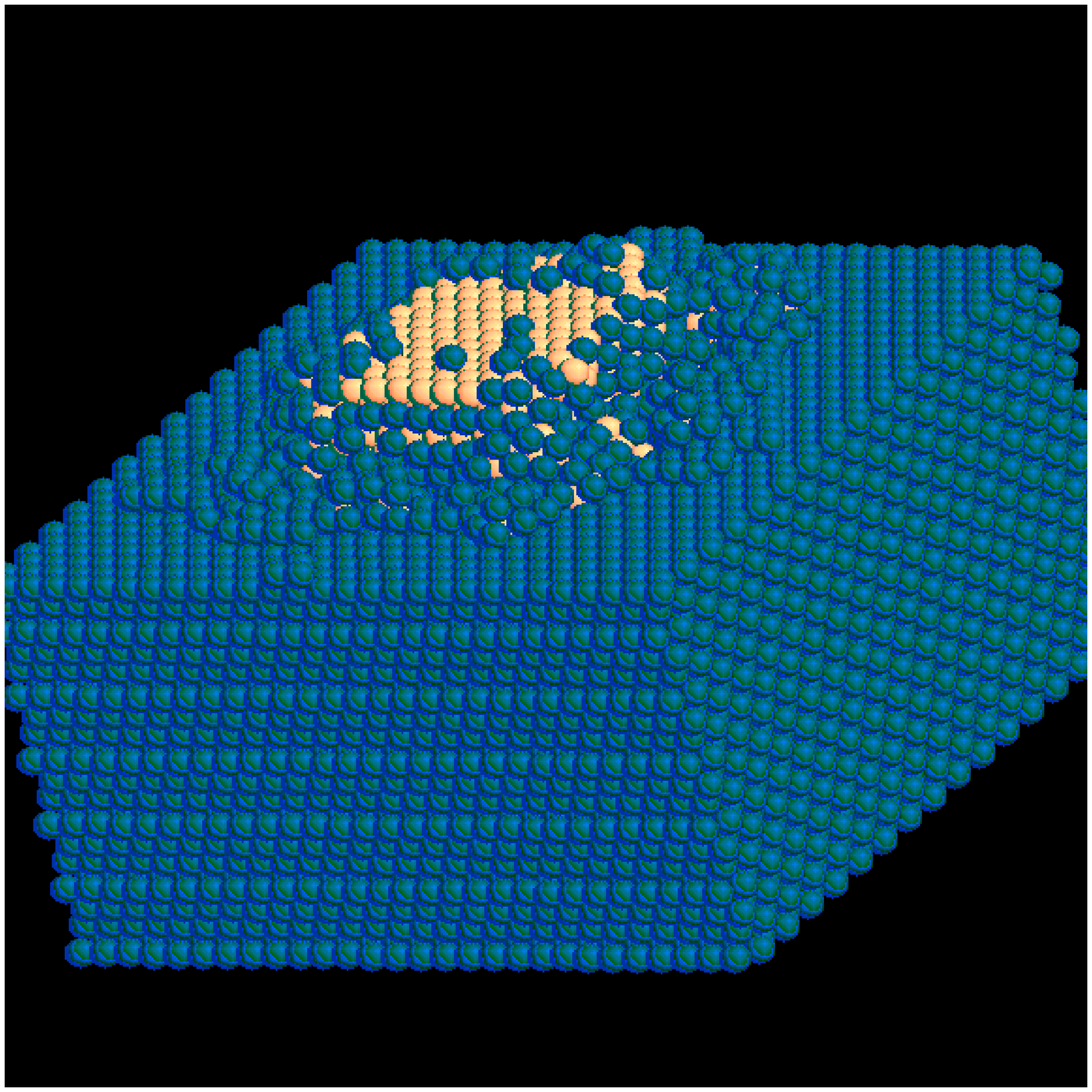}
\includegraphics*[height=3.cm,width=4.2cm,angle=0.]{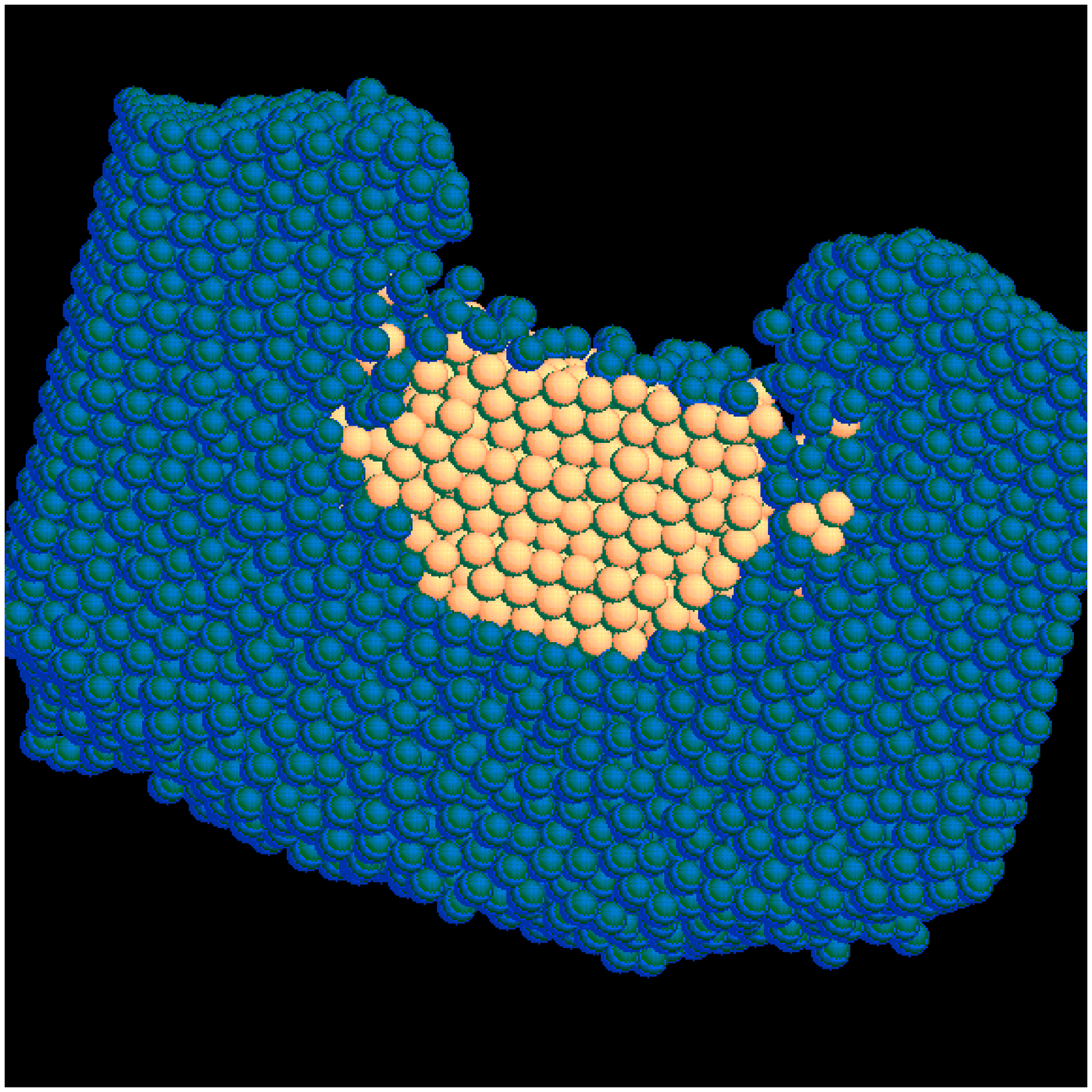}
\caption[]{
The cartoons of the simulation cells obtained after the impact of
Co cluster on Au(111) and Pt nanocluster on Al(111).
The atoms with lighter color are the cluster atoms.
From right to left:
{\em fig. 2a:} the top view of the simulation cell of Co cluster on Au(111) impact event
after 5 eV/atom impact.
{\em fig. 2b:} the burrowed (embedded) Pt cluster in Al (crossectional view, 1eV/atom impact
),
{\em fig. 2c:} the top view of the burrowed Pt cluster,
{\em fig. 2d:} the crossectional slab (cut in the middle of the cell) of the Al(111) substrate after the penetration of
the implanted Pt cluster (5 eV/atom impact energy).
Figs. 2a-2c have been plotted after few tens of ps the cluster impact.
In Fig. 2d the snapshot is plotted at $5$ ps.
}
\label{ptonal}
\end{center}
\end{figure}
which indicates that the
interpolation 
scheme is effective.
We assume that the fitted diatomic potential is transferable for
those cases when a cluster atom is embedded in the substrate.
Heteronuclear interactions can be reasonably well described
by dimer potentials during interfacial interactions while
pair potentials are less accurate in alloy phases where
the local number of crossinteractions is much larger than $1$.

 In Figs 1a-1c we show the {\em ab initio}
DFT potential energy curves and the fitted interpolated ones.
In most of the cases we could reach a fairly good agreement which
might increase the credibility of the employed approach.
We also carried out simulations for couple of other cluster/substrate pairs and for
which we use the simple interpolation scheme for the crosspotential.
Results for these pairs are shown in the discussion section.

\section{Results}


   The low-energy deposition ($1-5$ eV/atom) of nanoclusters of Pt on Al(111) and Ti leads to the unexpected
ultrafast cluster sinking and burrowing within few ps \cite{web}. 
The process can be considered as an athermal transient transport which takes place
in nearly $0$ K simulations within a ps at few eV/atom imapct energy.
We find a similar behavior for other cluster/substrate couples such as Cu/Al, Ni/Al, Ni/Ti or Co/Ti, etc.
Common feature of these couples is the large atomic size or mass anisotropy
(the host atoms are large and light and
the cluster atoms are smaller and heavy).

 Normally, most of the clusters should stop and fall apart and spread on the surface
or in the upper layers at this energy regime
(see e.g. in the review article \cite{Jensen} or in refs. \cite{cluster,Pratontep,Carroll}).
Actually, this feature of cluster impact can be used for thin film growth \cite{clustersatsurfaces,Jensen,cluster}.
In this article,
this has been demonstrated for the Co cluster deposition on Au(111) in Fig. 2a and can also
be seen at a web page \cite{web}.
In this case we find the fragmentation of the Co cluster upon impact on Au(111) at 5 eV/atom 
impact energy. We expect for most of the low-energy cluster deposition events 
with various substrates and clusters
the stopping of the cluster on the surface (with only weak penetration).
Even for higher impact energies we find no penetration of Co to Au (up to few tens of eV/atom).
At 5 eV/atom energy the Co nanoparticle falls apart and an intermixed nanodot
remains on the surface.

  Also, no burrowing occurs for the couples with reversed succession: e.g. the energetic deposition
of Ti cluster on Pt substrate leads to trapping on the surface with some fragmentation and/or spreading on the surface
(similar situation takes place for Al/Pt and Ti/Ni, etc.) \cite{web}.
Hence the ultrafast burrowing of clusters is asymmetric with respect to
the interchange of the atomic constituents.

 Clearly we find a different and anomalous situation for Pt cluster on Al(111) and
for the other couples under study (Pt/Ti, Ni/Ti, Cu/Al).
 Instead of the normal cluster low-energy impact behavior the Pt cluster keeps its integrity
during the impact and starts to burrow (penetrate) ballistically below the surface instead of
stopping at the surface.
 Below $1$ eV/atom the stopping or the partial penetration of the Pt clusters
has been found which implies that the burrowing has a potential energy barrier of $\sim 1$ eV/atom.

  Penetration of clusters upon impact has only been found until now at much higher impact energy
above $\sim 15$ eV/atom which takes place together with the fragmentation of the cluster \cite{Jensen,Pratontep,Carroll}.
Hence no burrowing process takes place and the higher implantation energy of
the clusters lead to the serious damage of the cluster and the substrate.
E.g. the amorphization of size-selected Au, Ag, and Si clusters have been found in graphite at few keV 
implantation energy \cite{Pratontep,Carroll}.
Our finding is also different from the cases found for Co cluster burrowing in Cu.
In this case the cluster penetration (CP) is relatively slow process and takes place in a ns time scale and is a thermally activated process \cite{burrowing}.

 It should also be stressed that the Pt cluster remains also intact 
at higher impact energies.
In Figs. 2b-2c the cartoons of the burrowing event is shown for the case
when the cluster is initialized by 1 eV/atom kinetic energy.
The sinking of the cluster takes place in few ps (ballistic process). 
Increasing the impact energy
up to 5 eV/atom the sinking process continues and the cluster penetrates through
a $\sim 50$ $\hbox{\AA}$ thick Al or Ti slab (Fig. 2d) in $10$ ps.
The cluster travels through the thin substrate slab
with a nearly constant speed ($\sim 1$ eV/atom) and remains
nearly intact when leaves the bottom of the thin slab.
The corresponding animations can be seen in a web page \cite{web}.
In Fig. 2d we show a snapshot of cluster implantation at 5 ps.
\begin{figure}[hbtp]
\begin{center}
\includegraphics*[height=5.5cm,width=6.5cm,angle=0.]{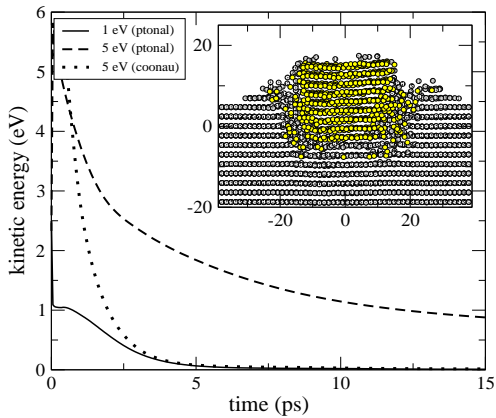}
\caption[]{
The average kinetic energy (eV/atom) of the penetrating Pt
cluster as a function of the simulation time (ps)
during complete penetration through the Al slab (the slab thickness is
$\sim 50$ $\hbox{\AA}$).
{\em Inset:} The crossectional view of the crossectional slab of the burrowed Pt cluster in Al at 1 eV/atom impact energy cut in the middle of the simulation cell.
}
\label{ekin}
\end{center}
\end{figure}

  Hence the stopping of the Pt cluster does not occur as it should be in most of
other cluster impact events (such in the case of Co cluster on Au or for Al cluster on Al). 
In the case of Co cluster on Au(111) we see the complete disintegration of the Co cluster on the
surface (Fig. 2a) even at 5 eV/atom impact energy.
At 5 eV/atom impact energy the Pt cluster penetrates far beyond the range of
the deposited energy (that is few $\hbox{\AA}$) into the substrate by an order of magnitude
larger implantation depth.
Therefore, simple collisional cascade effects can not explain the enhancement of
the vertical cluster mobility.
Therefore, there must be a specific mechanism which weakens stopping effects for the Pt/Al,
Pt/Ti for other couples.

 In Fig. ~\ref{ekin} the average (downward) kinetic energy of
the moving (burrowing) cluster per atom is also shown for a typical transient cluster burrowing (TCB) event at 5 eV/atom
initial kinetic energy. 
The cluster slows down to $\sim 1$ eV/atom
after the impact and continues moving through the Al slab.
Contrary to this, the Co cluster stops immediately after the impact and
the kinetic energy of the cluster drops sharply to zero at the same initial
impact energy.
In the case of 1 eV/atom impact energy we find burrowing for Pt on Al and no
further penetration is seen
which indicates that complete cluster penetration
has a barrier of few eV/atom.
The crossectional view of the embedded cluster is shown in the
inset Fig. ~\ref{ekin} at the end of the simulation which shows us
that the cluster remains largely intact during burrowing.
Also, even no partial cluster burrowing occurs below $\sim 0.5$ eV/atom impact energy.

\begin{figure}[hbtp]
\begin{center}
\includegraphics*[height=5.5cm,width=6.5cm,angle=0.]{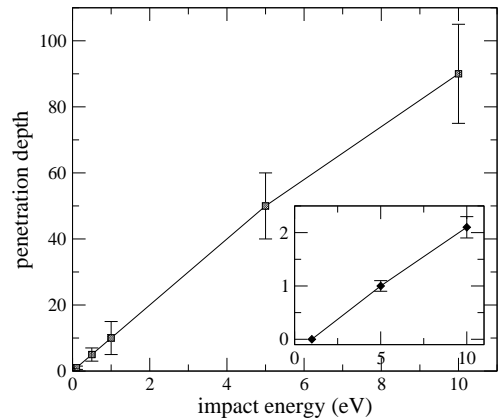}
\caption[]{
The penetration depth ($d$ in $\hbox{\AA}$) of various cluster impact events as a function of
the impact kinetic energy. Error bars denote standard deviations
obtained during various events (statistics).
{\em Inset:}
The average kinetic energy of penetration of cluster atoms (eV/atom) through the substrate as a function of
the impact kinetic energy (eV/atom).
}
\label{penet}
\end{center}
\end{figure}

 In Fig ~\ref{penet} we show the calculated penetration depth $d_p$ of clusters obtained
during various impact events making statistics as a function of the impact
energy up to $10$ eV/atom.
We find the nearly linear increase of $d_p$ with $E_{imp}$.
This is in accordance with the finding of Pratontep {\em et al.} \cite{Pratontep,Carroll}
as they found the nearly linear scaling of the implantation depth
of Au, Ag, and Si clusters into graphite as a function of the impact energy
in the keV impact energy regime.
We find this scaling relation at much lower impact
energy (few eV/atom regime).
In the Inset Fig ~\ref{penet} the linear scaling of the kinetic energy of penetration ($E_{pen}$, propagation)
can be seen as a function $E_{imp}$.
These features show us that both $d_p$ and $E_{pen}$ scales linearly with
$E_{imp}$.

\section{Discussion}

 Interestingly,
 the observed anomalous processes (transient burrowing,
the asymmetry of CP)
are only weakly sensitive to the choice of the heteronuclear potential.
We get similar results for the interpolated and for the fitted potentials.
At first sight this is surprising, however, we find that not the strength of
the crossinteraction built in
the crosspotential determines cluster mobility.
This is in accordance with recent findings in similar systems for atomic deposition
events \cite{Sule_JCP} and for ion-beam intermixing \cite{Sule_PRB05}.
The variation of parameter $\xi$ in Eq. (1) does not affect significantly the final results
in accordance with earlier findings during ion-mixing of the Ti/Pt bilayer \cite{Sule_PRB05}.
Parameter $\xi$ is proportional to the built in heat of mixing of the 
corresponding alloy phase
\cite{Sule_PRB05}. 
Atomistically, it determines the deepness of the potential energy well,
hence the strength of the
heteronuclear interaction.
The possible reason of this insensitivity to the strength of the crossinteraction is
that transient (ballistic) and athermal atomic transports are only weakly sensitive to pair-interactions
e.g. between colliding ballistic particles e.g. in a collisional cascade.
In other words, not chemical interdiffusion (chemical forces) and 
thermochemistry (heat of mixing) determine the magnitude of bulk cluster mobility
(and the implantation or penetration depth).
CP occurs e.g. in Pt/Al either for weakly repulsive and for attractive crosspotentials.

 \subsection{Mass-effect and cluster penetration}

  It turned out during comparing the penetration depths obtained for various cluster/substrate
pairs that atomic mass anisotropy, that is the atomic cluster to substrate
mass ratio ($\delta$) could seriously influence CP and could be responsible for the
apparent asymmetry in CP with respect to the interchange of the cluster and
substrate constituents.

 Since we find that atomic mass ratio could be an important ingredient
of CP we briefly summarize our recent results in which we find the
strong effect of mass anisotropy on interdiffusive behaviors of various
film/substrate systems \cite{Sule_PRB05}.
  The enhancement of atomic penetration (intermixing) has also been
reported recently of Pt in Ti substrate during ion-sputtering of Pt/Ti film/substrate
system \cite{Sule_JAP07}.
The asymmetry of interdiffusion in Pt/Ti with respect to the succession
of the film and substrate has also been found \cite{Sule_JAP07}.
In few other recent publication we also explained the enhanced atomic transport with
mass-effect \cite{Sule_PRB05,Sule_NIMB04b}
.

 It might be the case that similar mass ratio driven thermalization appears during CP.
Indeed, if we set in artificial mass isotropy in burrowing systems, we get
a much weaker cluster penetration.
Moreover, if we interchange artificially atomic masses between Al and Pt, we get no CP.
Hence we find that the inversion of mass-anisotropy in the system
hampers the enhancement of cluster mobility and CP.
Moreover,
if we interchange atomic masses in the Co/Au system, we
does get, CP.
In mass isotropic systems, in which no cluster penetration occurs, such as Ni/Cu or Co/Cu, ultrafast CP can also be induced
by setting in artificial mass anisotropy.
These simulation results strongly suggest that mass anisotropy could be 
a decisive parameter in cluster implantation.

  The role of mass effect in CP
 might be due to collisional cascade effects. When cluster impact occurs
the initialization of local melting could not appear with atomic mass ratio
$\delta=\frac{m_{cl}}{m_{subs}} \le 1$, where $m_{cl}$ and $m_{subs}$
are the atomic masses in the cluster and in the substrate.
This is simply due to backscattering effects when $\delta \le 1$:
light atoms stop at an interface composed of heavy atoms at low impact energies
in a similar way as it has been found for intermixing during the ion-bombardment
of various bilayers \cite{Sule_PRB05}.
In the case of $\delta \approx 1$, however, the dissipation of the impact energy
into the lattice is the most effective \cite{Sule_PRB05}, and no thermalization
or only a short lived collisional cascade takes place which does not allow CP.
However, when $\delta \ge 1$, local thermalization of the substrate will appear 
because heavier cluster atoms penetrate into the substrate top layer.

\begin{figure}[hbtp]
\begin{center}
\includegraphics*[height=6.cm,width=7cm,angle=0.]{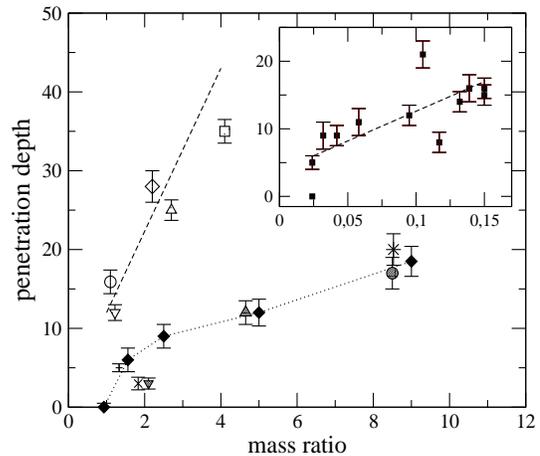}
\caption[]{
The penetration depth ($d_{pen}$ in $\hbox{\AA}$) of various cluster impact events ($2$ eV/atom) as a function of
mass ratio ($\delta$, $\delta \ge 1$). Error bars denote standard deviations
obtained during various events (statistics).
Filled diamonds correspond to values obtained for the Co/Cu couple.
The dotted line is for guiding eyes.
Plus, circle, star, cross, triangle down and triangle up denote Cu/Ti, Pt/Al,
Au/Al, Pt/Ag, Ti/Al and Au/Ag, respectively.
{\em Opened symbols}: Results are shown with opened
circle, diamond, triangle down, triangle up and square for Ni/Ti, Ni/Al, Co/Ti, Cu/Al and for Pt/Ti, respectively.
A linear fit has also been plotted with a dashed line to guide eyes.
{\em Inset:} The penetration depth ($\hbox{\AA}$)
as a function of the lattice mismatch $\epsilon$ ($\epsilon > 0$) for various
cluster/substrate couples which are the followings:
$\epsilon=0.024$, Cu/Co, $0.058$, Pt/Ti (iso),
$0.04$, Pt/Ag (iso), 
$0.10$, Ni/Pd (iso),
$0.11$, Cu/Al (iso),
$0.12$, Cu/Ag (iso),
$0.132$, Cu/Ti,
$0.14$, Ni/Ag (iso),
$0.15$, Co/Ti, $0.15$, Ni/Ti,
$0.15$, Co/Ti, $0.15$, Ni/Ti,
where (iso) denotes artificial mass isotropic
calculations.

}
\label{penet2}
\end{center}
\end{figure}

 The effect of $\delta$ on CP 
can be demonstrated more quantitatively if
we vary $\delta$ artificially e.g. in a nearly mass and atomic size isotropic system, such as Co/Cu and
plot the penetration depth as a function of $\delta$.
We use for simulations a standard Cleri-Rosato set of parameters \cite{CR}
and a fitted potential for the Cu-Co interaction \cite{Levanov}.
In principle we can get a universal plot, which holds more or less precisely for
any kind of cluster/substrate couples in which mass anisotropy
governs CP.
In Fig. ~\ref{penet2} we show 
with closed symbols those values
which are obtained
for those couples, which composed of atoms with
moderate lattice mismatch (e.g. in Cu/Ti, Pt/Al,
Au/Al, Pt/Ag, Ti/Al and Au/Ag). These systems give $d_{pen}$ values around the curve
obtained for Co/Cu.

 \subsection{The role of lattice missmatch}

 However, in those cases, in which both cluster to substrate lattice mismatch ($\epsilon=(a_s-a_{cl})/a_s > 0$, where
$a_s$ and $a_{cl}$ are lattice constants of the substrate and cluster
constituents)
and mass anisotropy are considerable
(Co/Ti, Ni/Al, Pt/Ti, Cu/Al) a deviation from the
curve of
Co/Cu
can be seen (opened symbols
and dashed fitted curve).
Hence we are faced with the splitting of the $d_{pen}$ vs. $\delta$ plot into two regimes.
These regimes can be understood as the separation of $\delta$ and $\epsilon$
dependent TCB processes.
This is simply due to the fact that one hardly can find systems in which
both $\delta$ and $\epsilon$ is robust.
In strongly mass-anisotropic couples with $\delta >> 1$ usually
$\epsilon$ is not that pronounced. When lattice mismatch is robust in such a way that the
cluster atom is the smaller
($\epsilon < 0$) mass anisotropy is not that robust as possible in other cases.
Hence the frequency of the appearance
of strongly mass anisotropic and lattice missmatched
($\delta, \epsilon$)
couples
among various possible cluster/substrate pairs is constrained by nature.

  In the nearly mass-isotropic Ni/Ti and Co/Ti only a smaller deviation can be seen,
in Fig. ~\ref{penet2}
although still a considerable penetration occurs in these size-mismatched couples.
In particular, the Co nanocluster burrows into the Ti phase temporarily completely 
and is ejected closer to the surface finally with $d_{penet} \approx 16 \pm 1.5$ $\hbox{\AA}$. 
However, no penetration takes place for the reversed cases (Ti/Ni, Ti/Co).

  {\em We argue than in this paper that atomic size mismatch (ASM, also we call it atomic size anisotropy) could be together with $\delta$ the basic parameters
which govern cluster mobility and TCB in the bulk.
}

\subsection{Burowing clusters: atomic size mismatch and cluster bulk mobility}

 Using
MD simulations it has been shown recently that the ratio of the cluster lattice parameter
to the substrate lattice constant (that is $\epsilon$) has significant effect on cluster diffusion
on the surface (interfacial incommensurability) \cite{Jensen,Deltour}.
In the rest of the paper we argue that transient cluster burrowing and bulk mobility could also depend on $\epsilon$.
The possible operation of a
$\epsilon$-dependent mechanism is supported by various simulations with mass isotropic
and lattice mismatched systems (e.g. Co/Ti, Ni/Ti, etc.) which also show up
transient cluster burrowing.


  Varying the elements in the cluster and substrate we can reach the
conclusion that ASM could also be a key parameter in TCB.
In particular, in the Ni/Cu system we get no TCB ($\epsilon \approx 0.02$), while if we replace Cu with Ti, TCB does occur
in Ni/Ti ($\epsilon \approx 0.15$). The former couple is a nearly mass and size isotropic system, while the latter one
is size anisotropic being a considerable atomic volume difference between Ni and Ti.
Also, no TCB occurs in the Ti/Ni system, when Ni is the substrate and Ti is the cluster.
The larger Ti atoms can not penetrate to the Ni phase.	
These situations support the ASM driven mechanism when $\delta \approx 1$.
  Also, an artificially set mass isotropy e.g. in Pt/Al ($\epsilon \approx 0.06$)
does not suppress completely CP, hence there must be another system parameter
which drives CP even in mass-anisotropic couples.
Co/Ti provides another mass isotropic and size anisotropic example in which
TCB occurs.
Finally, if we deposit Ti cluster on Al, no TCB occurs in accordance
with the expectations (size isotropic couple).
 
 These results are summarized in the Inset Fig ~\ref{penet2} in which
the penetration depths of various cluster/substrate couples at $2$ eV/atom
impact energy have been plotted against the cluster to substrate lattice mismatch
$\epsilon$.
In the mass anisotropic cases in order to exclude mass effect on CP
and to study purely the effect of lattice mismatch on CP
an artificial mass isotropy has been imposed (iso).
We assume that lattice mismatch and mass anisotropy have a nearly independent effect on 
CP, and their contributions to the total penetration depth $d_{pen}$ can be summed up
from individual terms: 
$d_{pen}=d_{pen}^{\delta}+d_{pen}^{\epsilon}$,
where 
$d_{pen}^{\delta}$ and $d_{pen}^{\epsilon}$ are the
mass anisotropy and atomic size mismatch induced contributions to transient
cluster penetration.
Hence during mass isotropic simulations 
$d_{pen} \approx d_{pen}^{\epsilon}$.
We find a nearly linear dependence for various couples which supports
the assumption that CP might depend on $\epsilon$.

  In few of the cases, however, deviation from the linear dependence has
also been found (Cu/Ag, Cu/Al).
We find the strongest transient burrowing process in iso-Cu/Al 
while
in iso-Cu/Ag the penetration depth drops below the fitted curve.
In these cases it could be that not only anisotropy parameters govern CP, it could be
that hidden (unknown) parameter(s) also contribute to CP.
For vanishing lattice missmatch ($\epsilon < 0.03$) $d_p$ vanishes abruptly if $\delta = 1$ (mass isotropy).

\subsection{Nonburrowing clusters}

 We summerize hereby the general properties of nonburrowing cluster/substrate pairs.
In general, in these couples cluster penetration and implantation can only
be induced at least with few tens of eV/atom impact energy.
For instance, in the Co/Cu system we find complete implantation at $\sim 50$ 
eV/atom energy, however, the cluster looses its integrity and even
the substrate experiences serious damage.
Hence, the technological applicability of this energy regime is questionable.

 As it has already been shown for many examples,
the interchange of atoms in a burrowing cluster/substrate couple
might lead to a nonburrowing couple.
In these couples $\epsilon < 0.0$ (negative lattice mismatch) and/or $\delta \le 1$
(mass isotropy, $\delta \approx 1$ or inversed mass anisotropy, $\delta < 1$).
The nonburrowing $\epsilon < 0.0$ couples are those A/B systems in which
A=Al, Ti, Ag, Au and B=Cu, Ni, Co, Pt.
Hence $\epsilon < 0.0$ together with $\delta < 1$ have a strong suppressive effect on TCB independently from each other.
The deposited cluster remains on the surface 
and $d_{pen} \approx 0$ when $\delta < 1$ and/or $\epsilon < 0.0$.
In these cases penetration does not occur even temporarily. In
many other mass-isotropic ($\delta \approx 1$) and size-anisotropic cases ($\epsilon > 0.0$) clusters first burrow deeply into the substrate
(e.g. in Co/Ti) and than are ejected to the surface or to the top layers.
%

 The thermally induced burrowing of Co clusters into Au(111) surface
has been studied recently \cite{Bulou}.
The experimentalists claim that Co clusters possess a strong
tendency to bury themselves into Au substrate \cite{Bulou}.
Since this process can only be activated above room temperature
annealing, we repeated our simulation for Co/Au with simulating
the annealing process and indeed the partial burrowing of the Co cluster
has been found at $450$ K, although the cluster does not keep
its integrity completely and sinks only few monolayer deep into the
substrate.
We conclude that in Co/Au the experimentally observed burrowing might not be
an athermal process (TCB) and could be similar to that of reported in the Co/Cu couple \cite{burrowing}.
 In the size-mismatched Co/Au system although $\epsilon = 0.13$, however, $\delta \ll 1$, hence Co clusters are stopped at
the Co/Au interface.
The stopping power of $\delta \ll 1$ is a stronger effect than the burrowing effect of lattice mismatch.
Hence we conclude from this that the effect of ASM can only be realized
when $\delta \ge 1$.
This has been demonstrated 
with an artificially inverted $\delta$ (the atomic masses of the cluster and substrate atoms have been interchanged) with which we can
induce ultrafast CP in Co/Au.
It is also clear that
no penetration appears in Au/Co, although $\delta > 1$.
In this case
this is due to the stopping power of $\epsilon < 0$.
Hence $\delta \ll 1$ and $\epsilon < 0$ suppress TCB.




\section{Conclusion}
 
  In conclusion, we explore the occurrence of 
transient nanocluster penetration (burrowing)
in various substrates using atomistic simulations
at low energy cluster impacts.
We point out that the cluster to substrate atomic mass and size anisotropy plays a significant role in the mass transport:
clusters with heavier and smaller atoms burrow into a substrate composed of lighter and larger atoms.
 The transient burrowing process is largely
insensitive to the strength of
cross-interaction (that is proportional to the heat of mixing) and is driven by
the atomic mass ratio and lattice mismatch.

 The deep penetration of metallic clusters into metallic substrates could allow the
preparation of metallic inclusions and burried nanostructures implanted into the substrate.
Beyond the possible technological application of cluster burrowing, the explored new
phenomenon could also be interesting in a theoretical point of view.
In particular, the understanding of transient cluster mobility in the bulk could contribute to the
advance of the emerging new field of anomalous diffusion (see refs. \cite{Sule_JCP,Sule_JAP07} and references therein).
These peculiar results provide new evidences to the new concept
that atomic transport could become anomalous in the nanoscale under not yet clearly established conditions \cite{Beke}.
Common feature of few of these processes the athermal characteristics 
(non-Arrhenius atomic transport)
and that the deposit-surface interaction is largely independent of
chemical forces \cite{Sule_PRB05,Sule_JCP}.

\section{acknowledgment}
This work is supported by the OTKA grants F037710
and K-68312
from the Hungarian Academy of Sciences.
We wish to thank to K. Nordlund, T. Michely and to M. Menyh\'ard 
for helpful discussions.
The help of the NKFP project of
3A/071/2004 is also acknowledged.

\end{document}